\documentclass{elsart}

\usepackage{epsfig}
\def\lsim{\hbox{ \raise.35ex\rlap{$<$}\lower.6ex\hbox{$\sim$}\ }}
\def\gsim{\hbox{ \raise.35ex\rlap{$>$}\lower.6ex\hbox{$\sim$}\ }}

\begin{document}
\begin{frontmatter}

\title{Why do we live in 3+1 dimensions?} 

\author[geneva]{R. Durrer}
\ead{ruth.durrer@physics.unige.ch}
\author[geneva]{M. Kunz}
\ead{martin.kunz@physics.unige.ch}
\author[geneva,athens]{M. Sakellariadou}
\ead{msakel@cc.uoa.gr,mairi@mpej.unige.ch}

\address[geneva]{Department of Theoretical
Physics, University of Geneva, 24 quai Ernest Ansermet, CH-1211 Geneva
4, Switzerland}
\address[athens]{Division of Astrophysics, Astronomy and
Mechanics, Department of Physics, University of Athens GR-15784
Zografos (Athens), Hellas}

\begin{abstract}
In the context of string theory we argue that higher
dimensional D$p$-branes unwind and evaporate so that we are left with 
D3-branes embedded
in a (9+1)-dimensional bulk. One of these D3-branes plays the r\^ole
of our Universe. Within this picture, the evaporation of the higher
dimensional D$p$-branes provides the entropy of our Universe.
\end{abstract}


\end{frontmatter}

\section{Introduction}
One of the open questions in modern cosmology is the dimensionality of
spacetime. Why do we live in a Universe of (3+1) dimensions? 
The question why the spatial dimension is not lower than 3
can be answered quite satisfactorily with a weak form of the anthropic
principle: if it were, then there would be no intelligent life around to
ask the question. But why is it not higher, e.g. 4, 5 or 42?

In this treatise we do not want to address the question in its full
generality, but we restrict ourselves to string theory. We assume that
a 10-dimensional super-string theory (type I, IIA or IIB; not a heterotic
type theory since it does not have D-branes)
be the correct description of the physical world. We also assume that
the theory lives on a 9-dimensional spatial torus, the tenth
dimension being time.

So far two possibilities  to reduce the 10 dimensions
from string theory to the 4  dimensions of the observed spacetime have
been under discussion.
\begin{itemize}
\item[{\bf (i)}] The Kaluza Klein approach, where the 6 extra-dimensions are
  rolled up in a small torus (or more generically a Calabi-Yau manifold)
  with a size given by the string scale  $\sqrt{\alpha'}$ which is
  much smaller than all scales probed in the laboratory. 
\item[{\bf (ii)}] The braneworld approach, where our observed Universe
  represents a 3-brane on which open strings can end (Dirichlet-brane or
  D-brane, see Ref.~\cite{Pol}). Since gauge charges 
  are attached to the ends of strings, gauge particles and fermions
  can propagate only along the 3-brane while gravitons (and
  dilatons, ...) which are closed string modes can move in the
  bulk. Since gravity has been probed only down to scales of about
  $ 0.1$mm, the dimensions of the bulk can much larger than the string scale.
\end{itemize}

In the braneworld context, the extra-dimensions can even be infinite,
if the geometry is non-trivial and they are warped~\cite{RSII}. Large
extra-dimensions can be employed to address the hierarchy
problem~\cite{Ark}. This and other attractive features have led to a
growing body of literature on braneworld models and their astrophysical
and cosmological consequences~\cite{Ark2,Roy}.

In both approaches, the number of large spatial dimensions is set equal to 3
just in order to agree with observations, but without physical motivation. The
first argument why the number of large spatial dimensions should be three has
been made within the Kaluza-Klein approach by Brandenberger and
Vafa~\cite{BV}. They have argued that, when allowing strings to wind around a
9-torus, they intersect and unwind only inside a 3-dimensional sub-manifold, so
that only three dimensions can grow large while the other 6 are held back by
strings wrapping around them. This hypothesis has been verified numerically by
Sakellariadou~\cite{MS}.

Here we argue within the braneworld approach. We claim that due to
intersections leading to reconnection and unwinding, all D$p$-branes 
of dimension $p>3$ disappear; they are unstable. One of the
stable 3-branes plays the r\^ole of our Universe. We focus our
discussion on type IIB string 
theory, but our results could also hold for any other version of
string theory which allows for 3-branes.

The picture we have in mind is that at very early times space was
potentially large and filled with D$p$-branes and 
${\bar {\mbox D}}p$-anti-branes of all possible dimensions $p$.
A brane differs from its anti-brane by possessing the opposite
Ramond-Ramond charge. Since the charge of a D$p$-brane corresponds to an
orientation, a ${\bar {\mbox D}}p$-anti-brane is a D$p$-brane rotated by
$\pi$. We postulate that at very early times, high, as well as low,
dimensional branes fill space. 

We investigate the intersections between D$p$-branes of various
dimensionality $p$, embedded in a $9+1$ dimensional toroidal
bulk. In Section 2, we first state the condition under which two
D$p$-branes intersect. For simplicity we disregard
intersection between two branes of different dimensionality.
We then argue that
brane intersections will eventually lead to evaporation of D$p$-branes
with $p>3$ into gravitons and dilatons,
and a system of D3-branes (and possibly D2- and D1-branes). One of the
D3-branes could become our
Universe. In the process of 'evaporation' of the higher dimensional branes,
 the entropy of the Universe increases.  We state our
conclusions as well as some remarks in Section 3.

\section{Brane Intersections}
We consider a uniform distribution of D$p$-branes
embedded in a higher dimensional bulk. We denote the spacetime
dimension of the bulk by $d$. We want to set the condition
for the intersection of two D$p$-branes.  A D$p$-brane is
$p$-dimensional with $p$ taking any even value in the IIA theory, any
odd value in the IIB theory, and the values 1, 5, and 9 in the type I
theory. We assume that branes at macroscopic distances do not
interact. Their intersection probability is then purely a question of
dimensionality, and the following statements are true  with
probability 1, \textit{i.e.} always except for branes which
accidentally have one or several parallel directions.  If $2p\geq
d-1$, then the two D$p$-branes  intersect at all times on an
intersection-manifold of dimension $2p-d+1$, while if  
$2p+1 = d-1$ then the two D$p$-branes intersect in an 'event',
\textit{i.e.} they  eventually intersect at
some time $t_c$ in a point. However, if 
$2p+1 < d-1$, the two D$p$-branes will generically never intersect. 
Thus, the condition for generic intersection of two D$p$-branes embedded in a
$d$-dimensional spacetime (the bulk) is
\begin{equation} \label{eq1}
2p+1 \geq d-1~.
\end{equation}

Let us consider the case of $d=10$. Then according to the condition
above, two D$p$-branes will never intersect if $p\leq 3$, but
 they will eventually collide provided $p\geq 4$.  

The simple dimensional condition~(\ref{eq1}) for the intersection of
generic D$p$-branes has also been mentioned in Refs~\cite{An1,An2}. 
But there, the authors do not conclude that this fact leads to the
disappearance of higher dimension branes. In Ref.~\cite{An1} they
argue that the density of higher-dimensional branes is exponentially
suppressed for entropic reasons and in 
Ref.~\cite{An2} they even say that due to the inter-brane potential, a
simple dimensional argument is not valid.

We now want to argue that intersecting branes are unstable and
eventually evaporate so that we are left
with D3-branes (and any permitted lower dimensional branes, D1-branes in
 type IIB theory). This is the main point of the present paper.
For our argument we need the following hypotheses:
\begin{itemize}
\item[{\bf (1)}]
We assume that the $9$ bulk coordinates are compactified on a torus.
Closed branes which do not wind around the torus shrink and
disappear emitting gravity waves (and dilatons or other closed
string modes). We call this process evaporation.
\item[{\bf (2)}]
 If a D$p$-brane intersects with another D$p$-brane on a
hypersurface of dimension $p-1$ the branes reconnect, this means
one side of the first brane reconnects with the other side of the
second brane and vice versa (see Figs.~1 and 2). In this way their
winding number is reduced until they finally do not wind anymore and
 thus can evaporate. 
\item[{\bf (3)}] If two D$p$-branes intersect on a manifold of dimension lower
than $p-1$ the open strings which switch between the branes lead to an
alignment/anti-alignment of the directions with the smallest respective
opening angle (see Fig.~3). This process continues until the intersection
manifold has dimension $p-1$ and the branes can reconnect and separate
again.
\item[{\bf (4)}]
We assume that the total winding number of all branes of a given
dimensionality vanishes.
\end{itemize}

\begin{figure}[ht]
\begin{center}\includegraphics[scale=.5]{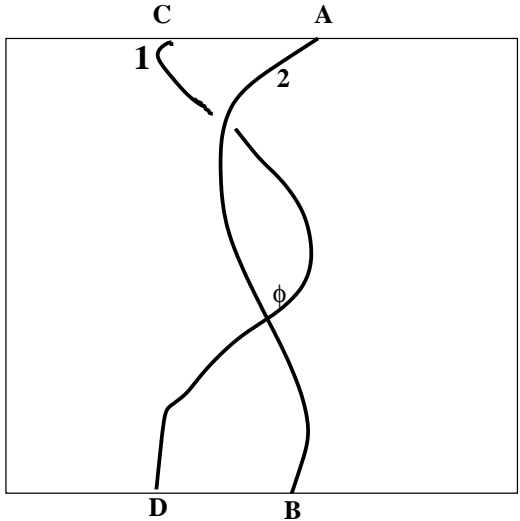}
\caption{The projection of two D$p$-branes, denoted by 1 and 2, which
  intersect along a $p-1$ dimensional manifold along dimensions
  omitted in this figure. They intersect in the point $\Phi$.
  We choose
  periodic boundary conditions. In a toroidal geometry, point A is
  identified with B, and point C with D.}
\end{center}
\end{figure}
\begin{figure}[ht]
\begin{center}\includegraphics[scale=.5]{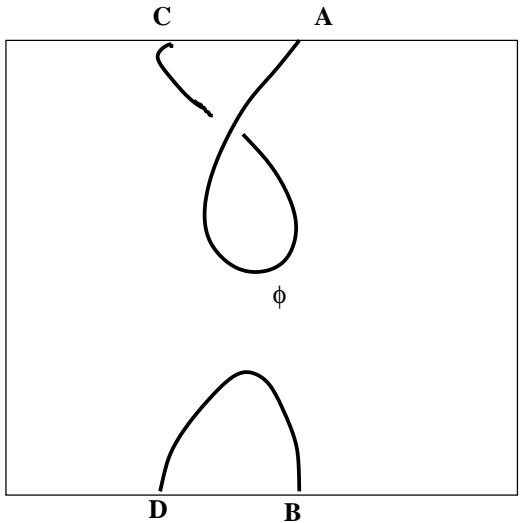}
\caption{The new D$p$-brane which results from the intersection of the
  two D-branes shown in Fig.~1. With respect to the directions shown
  in the figure, it no longer winds around the torus.}
\end{center}
\end{figure}

\begin{figure}[ht]
\begin{center}
\includegraphics[scale=1.]{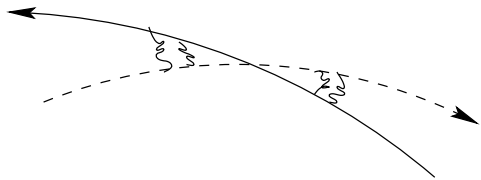}
\caption{Schematic representation of two intersecting branes and the
  open strings which are attached to both of them. They lead to
an anti-aligment of the two branes.}\label{fig3}
\end{center}
\end{figure}

Let us briefly comment on each of these hypotheses. Point~{\bf (1)}
 is quite natural. A simple entropy argument
 implies that a state of many gravitons is entropically favored
over a state with a brane. If it is not topologically forbidden,
 evaporation will therefore take place. This process  leads to
entropy production in the bulk.

Point~{\bf (2)} seems also unproblematic. It has been verified
numerically for Nambu-Goto strings in~\cite{MS}. More realistically,
the branes might have some reconnection probability $P<1$ but this
does not change our argument qualitatively. For intersecting
D-strings at an angle $\phi$, it has been shown
analytically in the low energy limit~\cite{HaNa}, 
that there is a tachyon mode which represents the instability to
reconnection. D$p$-branes which intersect in $p-1$ directions can be
reduced to this case by applying T-duality in the $p-1$ common directions.

Point~{\bf (3)} is a crucial assumption for our scenario to work. We sketch
here the reasons for which we expect that this assumption does hold.

If branes are parallel and have vanishing relative velocity,  some of
the supersymmetries are preserved.  In this case, the
Ramond-Ramond repulsion cancels exactly the gravitational and dilaton
attraction; the potential energy is zero. But in the general case, and even
more importantly in the case of several dynamically moving branes, one expects
that D$p$-branes are at general angles to each other, implying that all
supersymmetries are broken~\cite{Pol}. In this case, there will be an
angle-dependent (or equivalently, velocity-dependent) force
acting on the branes.  This force corresponds to fundamental strings
attached to both 
branes. For non-zero angles, and intersecting branes, these open strings
will be confined to the region near the intersection (see Fig.~\ref{fig3}).

At low energy, the interaction between branes can be described by a
well known interaction potential, which comes from the scattering
amplitude of open strings which
end on the two branes (which can also be viewed as closed strings
travelling from one brane to the other, the cylinder amplitude). 
In this case, the force between the branes is simply the 
gradient of the interaction potential. In Ref.~\cite{Pol} the potential
is given for the case of two D4-branes at some minimal separation $y$. 

In the simplest example of only one non-vanishing angle between the two
D4-branes, the lightest excitation has the energy (mass)~\cite{Pol} 
 \begin{equation}
m^2={y^2\over 4\pi^2\alpha'^2}-{\phi\over 2\pi\alpha'} ~~~~{\rm with}~~~~
0<\phi\leq \pi~,
\end{equation}
where $y$ is the closest separation between the four
branes. When the branes  come close, this mode becomes tachyonic
(negative $m^2$) indicating an instability. When $\phi=\pi$, the
branes are anti-aligned and form a D4-brane/anti D4-brane
configuration which will annihilate. But even if the branes are nearly
aligned, $\phi \ll \pi/2$, a tachyonic mode appears once their separation
$y$ is small enough, $y\ll \sqrt{\alpha'}$. The branes can lower their
energy by reconnection, which will eventually lead to unwinding. This
confirms again our point {\bf (2)}. In this case, the branes are
already aligned in 3 directions, i.e. they intersect on a 
$3(=4-1)$-dimensional sub-manifold and can thus reconnect.

\begin{figure}[ht]
\begin{center}
\includegraphics[scale=.65]{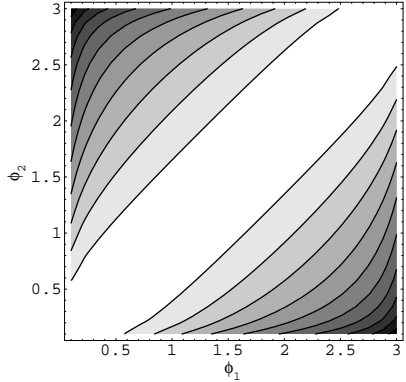}
\caption{The interaction potential $V(\phi_1,\phi_2)$ for two D4-branes
  which intersect on a plane and have two directions which are not
 aligned. The diagonals $\phi_1=\phi_2$ and $\phi_1=\pi-\phi_2$ are
 symmetry axes of the potential. The potential is exactly zero for
$\phi_1=\phi_2$ and negative everywhere else. A configuration initially in 
  one of the four quadrants will always move to the closest boundary
  of the plot, which corresponds to an alignment or anti-alignment in one
  of the directions.}
\label{pot}
\end{center}
\end{figure}

The potential for the more interesting case of two angles is shown in
Fig.~\ref{pot}. As one can see, a D4-brane initially at angles
$\phi_1,\phi_2$ prefers to align  the smaller of the two angles (or
anti-align the one which is closer to $\pi$). Then, if
the brane distance $y$ is small enough, reconnection can take place.

We expect this alignment to proceed locally, and in a causal way.
Once there is a region where the branes intersect 
 in $p-1$ dimensions, they can
reconnect there and we expect that a 'wave of reconnection' appears
which moves outward until finally the branes have entirely reconnected.
We hope to give a detailed description of this picture in~\cite{prep}.

The lightest mass of two D4-branes at four arbitrary angles, 
 $0< \phi_i \le\pi$, depends on the largest angle. If this is e.g. $\phi_1$,
the lightest mass is 

\begin{equation}
m^2_1={y^2\over 4\pi^2\alpha'^2}+{\phi_4+\phi_3+\phi_2\over
  2\pi\alpha'} - {\phi_1\over 2\pi\alpha'}~. 
 \end{equation}
Hence the energy can be lowered by aligning the angles $\phi_2$ to
$\phi_4$. As soon as $y$ and  $\phi_2$ to $\phi_4$ are sufficiently
small, a tachyonic mode appears which indicates an instability which
probably leads to reconnection. Of course a corresponding mode
exists for each angle $\phi_i$ by symmetry reasons, however, the
lowest mass mode is the one determined by the largest angle.

The example discussed here, two D4-branes, is not relevant for
type IIB string theory and we have chosen it, because it is
treated in detail in Ref.~\cite{Pol}. Nevertheless, from the
generality of the interaction potential it is clear, that the situation
will be very similar for D5-branes. Apart from having one more angle
which is readily incorporated,
the main difference is that the minimal distance between
D5-branes in 9-dimensional space vanishes. They generically intersect along
a line, except in the special case when they are parallel in at least
two directions. 
By applying T-duality along the intersecting direction we end up with
D4-branes which intersect in a point, $y=0$.
The general expression for the potential given
in  Ref.~\cite{Pol} diverges for vanishing brane distance and we thus have
to make a more thorough analysis in this case which we postpone to
later work~\cite{prep}. 

A detailed investigation of two D2-branes intersecting in a point at
two angles is given in Ref.~\cite{Na}. There it
is found that tachyon condensation leads to 'local diffusion' of the
two D2-branes near the intersection point. We shall argue that to the
next order, tachyon condensation leads to a reduction of the smaller
of the two angles, thereby rendering the two branes more parallel
along this direction~\cite{prep}, leading finally to a region where we
have $p-1$ nearly parallel directions and the branes can start to reconnect.

Point {\bf (4)} is probably not very important. If it is not satisfied, then
for topological reasons some D$p$-branes may remain even if $p>3$,
but nevertheless they would be much rarer than D3-branes.

Our scenario takes place within the framework of ten-dimensional type
IIB supersymmetric string theory. Hence the number of spatial dimensions of the
bulk is $d-1=9$, and the possible dimensionality of the D$p$-branes is
$p=1, 3, 5, 7$ or 9. We consider an initial state where the bulk is
filled with a 'gas' of all allowed  D$p$-branes. Assuming the
correctness of our hypotheses  {\bf (1)},   {\bf (2)},  {\bf (3)}, and 
 {\bf (4)}, after some time only D3- and D1-branes survive:

D9-branes are space-filling and the bulk coincides
with their world-volume. For a D9-brane there is no
partition of spacetime into Neumann and Dirichlet directions.  Since
D9-branes overlap entirely, they can immediately reconnect in a way
that the winding number of each of them vanishes and thus evaporate.

Two D7-branes, or two D5-branes, will always intersect on manifolds
of dimension 5 and 1 respectively. The D7-branes then have to align
along one
direction before they can reconnect and eventually unwind. The 
D5-branes have to align three directions before they can  reconnect.

We therefore expect first the D9-branes to evaporate, then the D7- and
the D5-branes last. At the end we are left with D3-branes and
D1-branes and a background of closed string modes (gravitons and
dilatons) in the bulk. 

Neglecting the interaction of branes with different dimensionality is
probably not a very good approximation. But since D3-branes
generically only intersect with D7- and D9-branes, as soon as those
have unwound and evaporated, the D3-branes are no longer affected
and survive.

\section {Remarks and conclusions}
We have addressed the question of why we live in (3+1)-dimensions in
the framework of braneworlds, where the standard model of
strong and electroweak interactions is described by open string
modes ending on branes, while gravitons and dilatons are the low
energy closed string modes in the bulk.  We 
consider IIB string theory and we claim that all D$p$-branes with $p$
greater than 3 disappear through brane collisions, emitting
fundamental string loops. After some time we are left with D$p$-branes
of 1 or 3 dimensions and a collection of closed strings in the bulk. 

It is interesting to note that the collision and evaporation of higher
dimensional D$p$-branes generates entropy by 
populating the bulk with gravitons and dilatons. If their density 
 is sufficiently high, we expect them to thermalize. 
These bulk modes also interact with the 3-brane representing our
universe and can be converted there into a thermal bath of all modes
living on the brane. This might explain the
entropy of our Universe. 

The mechanism discussed in this paper does not
address the question of how gravity gets localised to the brane.

\ack
It is a pleasure to thank Riccardo Sturani and Andr\'{e} Lukas for useful discussions.
This work is partially supported by the Swiss National Science
Foundation. MS acknowledges support from CMBNET.


\end{document}